\documentclass[conference]{IEEEtran}
\IEEEoverridecommandlockouts
\usepackage{cite}
\usepackage{amsmath}
\usepackage{amsmath,amssymb,amsfonts}
\usepackage{algorithmic}
\usepackage{graphicx}
\usepackage{textcomp}
\usepackage{xcolor}
\usepackage{subfigure}
\usepackage{multirow}
\usepackage{float}
\usepackage{caption}
\usepackage{mathtools, cuted}
\usepackage[export]{adjustbox}
\usepackage{booktabs}

\DeclarePairedDelimiter{\diagfences}{(}{)}
\newcommand{\C}{\mathbb{C}}
\newcommand{\diag}{\operatorname{diag}\diagfences}
\def\BibTeX{{\rm B\kern-.05em{\sc i\kern-.025em b}\kern-.08em
    T\kern-.1667em\lower.7ex\hbox{E}\kern-.125emX}}

\newcounter{tempEquationCounter} 
\newcounter{thisEquationNumber}
\newenvironment{floatEq}
{\setcounter{thisEquationNumber}{\value{equation}}\addtocounter{equation}{1}% record equation as happened and remember number
\begin{figure*}[!t]% float following equation across columns
\normalsize\setcounter{tempEquationCounter}{\value{equation}}% record current equation number in floated location
\setcounter{equation}{\value{thisEquationNumber}}% use previous equation number
}
{\setcounter{equation}{\value{tempEquationCounter}}% set back to equation number in floated location
\hrulefill\vspace*{4pt}% add a horizontal rule separator
\end{figure*}% end float environment

}

\begin{document}

\title{System Performance Insights into Design of RIS-assisted Smart Radio Environments for 6G
}

%\title{Toward Design of RIS-assisted Smart Radio Environments for 6G: A Performance Perspective}

\author{\IEEEauthorblockN{Mehdi Toumi, Adnan Aijaz}
\IEEEauthorblockA{\text{Bristol Research and Innovation Laboratory, } 
\text{Toshiba Europe Ltd., Bristol, United Kingdom  }\\
%Bristol, United Kingdom \\
mehdi.toumi@toshiba-bril.com, adnan.aijaz@toshiba-bril.com}
\vspace{-1.5em}
%\and
%\IEEEauthorblockN{Adnan Aijaz}
%\IEEEauthorblockA{\text{Bristol Research and Innovation Laboratory} \\
%\text{Toshiba Europe Ltd., Bristol, United Kingdom  }\\
%%Bristol, United Kingdom \\
%adnan.aijaz@toshiba-bril.com}

}    

\maketitle

\begin{abstract}
In state-of-the-art wireless networks, the radio environment is beyond the control of the operators which leads to several issues. For example, signal attenuation limits radio connectivity, multi-path propagation results in fading phenomena and reflections/refractions from large objects are the main sources of uncontrollable interference. To overcome these issues, a new technology referred to as reconfigurable intelligent surfaces (RISs) has been brought to light.  RISs, whose interactions with the electromagnetic waves are reconfigurable, are at the heart of the \emph{smart radio environments} for beyond 5G (or 6G) wireless systems. Research on design of smart radio environments is still in infancy. To this end, this paper provides performance insights for design of RIS-assisted smart radio environments. By extending a recent system-level simulator and through extensive simulations, it investigates the impact of positioning/placement of RISs, the number of reflecting elements and the tilt/rotation of RISs in both single and multi-RIS environments. Results reveal that these aspects are crucial to achieving capacity and reliability enhancements in smart radio environments.

\end{abstract}


\section{Introduction}

% wireless \cite{aijaz2018tactile}.

Current technologies assisting the Physical (PHY) layer in  wireless communications consist mainly of multiple-input multiple-output (MIMO) / massive MIMO beamforming, relaying and backscattering. These technologies are used for the purpose of tackling the limitations of scatters from large scale buildings and small-sized objects. The reason behind this vulnerability is due to the use of sub-6 GHz frequency bands which will soon not be able to meet the growing data rate requirements and will necessitate migration to mmWave and higher frequencies. Furthermore, legacy wireless communications lacks control over the wave propagation environment. The new wireless environment is envisioned within an architectural platform where the propagation of radio waves is controllable and programmable. 
This concept goes by \emph{smart radio environment} (SRE) \cite{9140329} and consists of intentionally and deterministically controlling the signal propagation in the radio environment.
%%of adding an optimization process during the propagation between the transmitter (Tx) and the receiver (Rx).  
The key technology underpinning SREs is reconfigurable intelligent surfaces (RISs) which have also been called metasurfaces, large intelligent surfaces (LISs) and intelligent reflective surfaces (IRSs). 
RISs consist of many anomalous controllable reflecting electromagnetic surfaces capable of changing the phase of the landing waves. 
It is a nearly passive new technology that, conversely to currently employed technologies, does not require signal processing nor signal amplification processes. The simple act of reflecting the waves from its surface spares the signal from additional noise. Moreover, it operates in full band spectrum and is easy and cheap to deploy.

\subsection{\textcolor{black}{Related Work}}

Ongoing research activities are investigating various aspects of RIS-assisted communication. 
Kudathanthirige \emph{et al.} \cite{kudathanthirige2020performance} conducted an analysis of achievable data rate and average symbol error probability for RIS-assisted communication between a Tx and a Rx. The accuracy of closed-form expressions has been shown to improve as the number of reflecting elements on the RIS increases. Hu and Rusek \cite{hu2020spherical} investigated deployment of RISs in three-dimensional (3D) as spherical surfaces which provide various advantages over their planar counterparts such as higher coverage and ease of deployment and placement. 
Ntontin \emph{et al.} \cite{ntontin2019reconfigurable} and  Bj{\"o}rnson \emph{et al.} \cite{bjornson2019intelligent} conducted thorough comparison of RIS and relaying technologies. Both studies conclude that RIS-assisted communication achieves better performance than relaying-assisted communication for large number of antenna elements. Moreover, it easier to manipulate impinging signals at an electromagnetic level than in the digital domain as in the case of decode and forward relays. Boulogeorgos and Alexiou \cite{perf_RIS} conducted a performance comparison of RISs and amplify-and-forward (AF) relaying systems. The authors show that RIS-assisted wireless
systems outperform  AF-relaying ones in terms of outage probability and ergodic capacity. 
Comparative studies of RIS and massive MIMO technologies have been the focus of some recent studies. The work of Dardari \cite{dardari2020communicating} shows that RIS offers more degrees-of-freedom (DoF) which is particularly beneficial for mmWave communication. Bj{\"o}rnson and Sanguinetti  \cite{bjornson2020power} compared power scaling laws and near-field behavior of the two technologies. The authors prove that an RIS-assisted setup cannot achieve a better signal-to-noise ratio (SNR) than an equal-sized massive MIMO setup. 
Some recent studies (e.g., \cite{he2020large} and \cite{he2020adaptive})  have also explored the use of RIS as a tool for positioning/localization.

%\textcolor{red}{Need to be revised} Another comparison between RIS and massive MIMOs are present in \cite{dardari2020communicating} and \cite{bjornson2020power} where Dardari in \cite{dardari2020communicating} and Bj{\"o}rnson and Sanguinetti in \cite{bjornson2020power} prove that RIS offers more degrees of freedom in the case of non-LOS channels or mmWave communications and in a general way, poor scattering-characterized-environments. 
%As from sensing and localization standpoint, He \emph{et al} \cite{he2020large} investigated RIS as a tool for objects localization and tracking in a mmWave MIMO network. The paper proves that RIS is beneficial for its sensing performance thanks to the focused transmitted signal or beam into the direction of the intended Rx. The performance of RIS is also proved to increase with larger number of reflecting elements of the intelligent surface. Not only RIS provides more accurate positioning but is also provides a high data rate as seen in \cite{he2020adaptive} where He \emph{et al} explore RIS communication and localization performance. 

\subsection{\textcolor{black}{Contributions and Outline}}

The objective of this paper is to provide system performance insights for design of RIS-assisted SREs. Existing studies evaluating the performance of RIS-assisted communication are mostly focused on theoretical frameworks and independently treat the impact of different parameters for a SRE. We adopt a simulation-driven approach and conduct a holistic evaluation of capacity and reliability aspects by simultaneously considering different parameters. We also evaluate system performance under single and multiple RIS-assisted SREs. 
%Throughout all the currently investigated RIS paradigms, many aspects of RIS are being explored from number of reflecting elements to number of RISs and their positions. However, these parameters are being explored independently from each other unlike what we did in this paper. In addition to that, we explored reliability in respect to the previously mentioned parameters. We also served from existing literature to simulate a multi-user environment and showcase RISs’ reflecting elements’ allocation topic.
To this end, our key contributions are summarized as follows.

\begin{itemize}
\item %Smart Radio Environment is a the (focus) of RIS. 
The positioning of RISs is a crucial factor for  RIS-assisted SREs. We prove that its closeness to the endpoints or its placement above the transmitter level boosts system performance. More importantly, it improves drastically in a multiple RIS environment. 
\item A key RIS application involves sensing and localization. In this paper, we explore the tilt and rotation aspects of RISs and investigate the system performance. Our simulations demonstrate performance enhancements when the RIS focuses the reflected beam toward the receivers.
\item The number of reflecting elements on a single RIS is another determinant  factor in the system performance. We prove that the capacity and reliability increase considerably for larger number of reflecting  elements. The channel shows a very slight improvement for small number of elements. However, multi-RIS configuration curtails performance issue of reduced number of reflecting elements. This highlights a trade-off in the design aspect of RISs.

\end{itemize}

The rest of the paper is structured as follows. Section \ref{Preliminaries} introduces the adopted simulator and our key enhancements. In Section \ref{sysmod} we provide a system model for RIS-assisted SRE. Section \ref{simulations} covers our performance evaluation of RIS-assisted SREs. Key insights, concluding remarks, and future work directions are provided in Section \ref{conclusion}.

%in section \ref{sysmod}, we introduce the system model as well as the channel model used in throughout this paper. Simulations and discussions follow that in section \ref{simulations} where we explore in depths the optimal placement of RIS and the performance of RIS for different number of reflecting elements in both single-RIS configuration at first followed by multiple-RIS assisted environment in a second phase. We finish this section by exploring the performance of RIS for multiple users scenario. Finally, in section \ref{conclusion}, we conclude on the paper providing a summary of the key findings and highlighting the road ahead in our RIS quest.

%and bringing an added value to the radio environment.

\section{SimRIS Simulator and Enhancements}\label{Preliminaries}
%Ongoing research is focusing on one-Reconfigurable Intelligent Surface based communications. SimRIS is one of the first simulations made about this new technology showing capacity improvements in a one surface assisted environment versus RIS-free environment. It also gave an overview about the improvement of capacity as a function of the position of the RIS \cite{basarsimris}. The improvements provided in this paper cover multiple-RIS assisted environment scenario and for which we have explored many parameters affecting this type of communications. Added to that, these parameters are explored in a multiple-RIS environment. Our simulations prove how advantageous the multiple RIS scenario is in developing a better communication. Not only it is improving drastically the communication environment but it is also more realistic. The deployment of such technology is envisioned over the  facades  of  buildings,the  windows of  homes or even on the underground walls. All of which would offer an opportunity for a Smart Radio Environment in a smart city.

%State of the art research is investigating many aspects of RISs. That includes the channel modelling and exploration of RISs’ performance by simulatingRIS-assisted environment which is the case in 
Our performance evaluation is based on a state-of-the-art simulator for RIS-assisted communication, i.e., 
SimRIS \cite{basar2020indoor}, which is an open-source channel simulator. It  models RIS-empowered radio environments over two operating frequencies (28 GHz -- 73 GHz) where for a given number of reflecting elements N, the user positions the N-element RIS in `xz' or `yz' plane in an indoor or outdoor environment. The channel responses h, g, \(h_{SISO}\) for the transmitter-to-RIS (Tx-RIS), RIS-to-reciever (RIS-Rx), transmitter-to-receiver (Tx-Rx) links respectively are given by the simulator. Nevertheless, in this paper, we compute our own RIS matrix (final result in \eqref{eq9}) that represents RIS contribution in the radio environment.
On top of that, we re-arranged the channel links to investigate the possibility to tilt and rotate the intelligent surface as can be seen in  \eqref{eq12}. We also aimed at exploring the influence of number of elements (N) dependently on number of RISs (M) deployed in a multi-RIS assisted environment \cite{yildirim2020modeling}. On that front, we addressed a trade-off between these two parameters. This could be a significant steppingstone for the physical conception of RISs. Another element of simulation is the multi-user scenario where we allocate certain reflecting elements from a single surface to each user. 
%This opens the gate for future research work that treats the optimal number of allocated elements to each user assuring stable connectivity.    

\section{System Model} \label{sysmod}

  \begin{figure}
  \centering
  \includegraphics[width=0.4\textwidth, center]{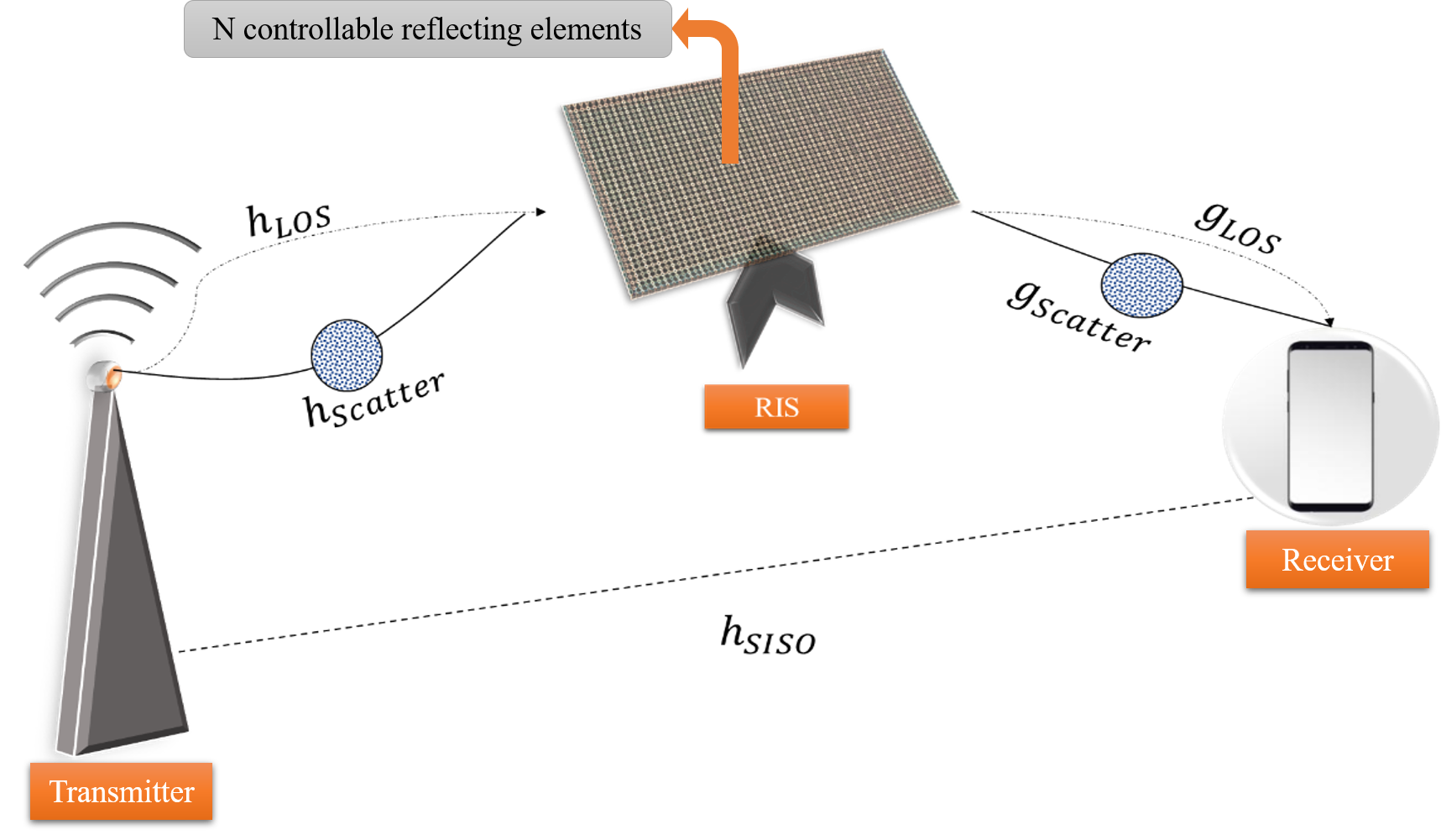}
  \caption{System model for RIS-assisted radio environment. }
  \label{f1}
  \vspace{-0.5cm}
  \end{figure}

Our system model for the RIS-assisted smart radio environment in Fig. \ref{f1} consists of a Tx, a Rx and an RIS composed of N controllable reflecting elements. If we consider h as the Tx-RIS channel response and g as the RIS-Rx channel response, we have \(h=h_{Scatter}+h_{LOS}\) and \(g=g_{Scatter}+g_{LOS}\) where \(h_{LOS}\) and \(g_{LOS}\) are line of sight (LOS) links. \(h_{Scatter}\) and \(g_{Scatter}\) are links due to scatterers. 
It has been discussed in \cite{basar2020indoor} that the received discrete time baseband signal can be written as:
\begin{equation}
y=\left(g^{T} \times \phi \times h\right) \cdot x
\end{equation}
where h\(\in \C^{N \times 1}\) and g\(\in \C^{N \times 1}\) (N\(\times\)1 complex matrices) and \(\phi\) is the matrix of RIS elements' response. However, in case of the presence of a direct link between Tx and Rx, it can be included in the expression above as follows \cite{basar2020indoor}: 
%$$
%y=\left(g^{T} * \phi \cdot h+h_{T_x-R_x}\right) \cdot x
%$$
\begin{equation}
y=\left(g^{T} \times \phi \times h+h_{T_x-R_x}\right) \cdot x
\label{eq2}
\end{equation}
%\textcolor{red}{(Comment: Use \(\times\) instead of asterisk)} 
In the same way the direct link between \(T_x\) and \(R_x\) is incorporated into the model, originating from the equation 2, we are suggesting that the received signal in case of adding a second RIS can be expressed as:
%$$
%y=\left(g_{1}^{T} * \phi_{1} * h_{1}+g_{2}^{T} \cdot \phi_{2} \cdot %h_{2}+h_{T_x-R_x}\right) \cdot x
%$$
\begin{equation}
y=\left(g_{1}^{T} \times \phi_{1} \times h_{1}+g_{2}^{T} \times \phi_{2} \times h_{2}+h_{T_x-R_x}\right) \cdot x
\end{equation}

In a more general way, for an M-RIS assisted system, the expression of the received signal can be written as:
%$$
%y=\left(\sum_{M} g_{M}^{T} * \phi_{M} * h_{M}+h_{T_x-R_x}\right) \cdot x
%$$

\begin{equation}
y=\left(\sum_{M} g_{M}^{T} \times \phi_{M} \times h_{M}+h_{T_x-R_x}\right) \cdot x
\label{eq4}
\end{equation}
 
The following content in this subsection is a summary of RIS channel response introduced in \cite{basar2020indoor}. We consider \(C\) the number of clusters and for each cluster c there are \(S_c\) scatterer.
The channel coefficients \((h)\) can be calculated as follows: 

%\begin{equation}
%\begin{split}
%h=\gamma \sum_{c=1}^{C} \sum_{s=1}^{s_{c}} \beta_{c, s} \sqrt{G_{e}\left(\theta_{c, s}^{R I S}\right) L_{c, s}^{R I S}} a\left(\phi_{c, s}^{R I S}, \theta_{c, s}^{R I S}\right)\\+h_{L O S}
%\end{split}
%\end{equation} 
\begin{equation}
h=\gamma \sum_{c=1}^{C} \sum_{s=1}^{s_{c}} \beta_{c, s} \sqrt{G_{e}\left(\theta_{c, s}^{R I S}\right) L_{c, s}^{R I S}} a\left(\phi_{c, s}^{R I S}, \theta_{c, s}^{R I S}\right)+h_{L O S}
\end{equation} 
where the normalization factor is
\(
\gamma=\sqrt{\frac{1}{\sum_{c} S_{c}}}
\),
\(\beta_{c, s}\) is the path gain following a complex normal distribution \(\mathrm{\mathcal{CN}(0,\,1)}\) and
\(L_{c, s}^{R I S} \quad\) is the path attenuation of the propagation in the \((\mathrm{c}, \mathrm{s})\) scatter path.
\(G_{e}\left(\theta_{C, S}^{R I S}\right) \quad\) represents the RIS gain for the \((\mathrm{c}, \mathrm{s})\) th scatterer and the RIS array response vectors is \(a\left(\phi_{c, s}^{R I S}, \theta_{c, s}^{R I S}\right) \in \mathrm{C}^{\mathrm{N} \times 1}\) which is a function of the Azimuth angle (RIS arrival) \(\phi_{c, s}^{R I S}\) and the Elevation angle (RIS arrival) \(\theta_{C, S}^{R I S} \quad\).
On the other hand, \(\mathbf{h}_{\mathrm{LOS}}\) can be calculated as:

\begin{equation}
h_{L O S}=I_{h\left(d_{T-R I S}\right)} \sqrt{G_{e}\left(\theta_{L O S}^{R I S}\right) L_{L O S}^{T x-R I S}} e^{j \eta} a\left(\phi_{L O S}^{R I S}, \theta_{L O S}^{R I S}\right)
\end{equation}
\hfill \break
where \(I_{h(d_{T-R I S})} \in\{0,1\}\quad \sim B(1, p)\) is a random variable that determines whether a LOS link exists for Tx-RIS path separated with \(d_{T-R I S}\). \(G_{e}\left(\theta_{L O S}^{R I S}\right) \quad\) is the RIS gain from the direction of LOS. As for 
\(L_{L O S}^{T x-R I S}\), it represents the attenuation of the LOS link in Tx-RIS path.
\(\eta \sim U[0,2 \pi] \quad\) is a random phase term and \(a\left(\phi_{L O S}^{R I S}, \theta_{L O S}^{R I S}\right) \quad\) is the RIS array response vector. \\
Second link in the RIS-assited communication path is g, which characterizes the RIS to Rx channel coefficients. g is evaluated through  \eqref{eqg} below.

\begin{equation}
g=\sqrt{G_{e}\left(\theta_{R x}^{R I S}\right) L_{L O S}^{R I S-R}}.e^{j \eta} a\left(\phi_{R x}^{R I S}, \theta_{R x}^{R I S}\right)
\label{eqg}
\end{equation}
where
\(a\left(\phi_{R x}^{R I S}, \theta_{R x}^{R I S}\right) \quad\) is the RIS array response vector \((\mathrm{Rx})\) and \(G_{e}\left(\theta_{R x}^{R I S}\right) \quad\) is the gain of RIS element in the direction of \(\mathrm{Rx}\). The attenuation of LOS RIS-Rx channel is denoted by \(L_{L O S}^{R I S-R}\) and  \(\eta \sim U[0,2 \pi]\) represents a random phase term. \\ 

Vis-à-vis the channel between Tx and Rx, it is represented in \(h_{SISO}\) and computed through the following equation: 
\begin{equation}
h_{\text {SISO }}=\gamma \sum_{1<c<C} \sum_{1<s<S_{c}} \beta_{c, s} e^{j \eta e} \sqrt{L_{c, S}^{\text {SISO }}}+h_{L O S}
\label{eqhsiso}
\end{equation}
First, we assume the use of single-input single-output (SISO) endpoints. Then we make the assumption that the RIS and Rx share the same clusters in Tx-RIS and Tx-Rx links, which results in the same \(\gamma\), C, Sc, \(\beta_{c,s} \) and finally the LOS component \(h_{LOS}\).
In  \eqref{eqhsiso}, \(L_{c, S}^{S I S O}\) represents the attenuation in path of Tx-Scatterer-Rx and \(\eta e\) is the phase excess due to the difference between the distance separating Tx-RIS and Tx-Rx through the same scatters. If we consider the scatterer (c.s) and its distances from RIS and from the Rx respectively denoted by \(b_{c, s}\) and \(\hat{b}_{c, s},\) we find:
\(\eta e=k\left(b_{c, s}-\hat{b}_{c, s}\right)\).

\subsection{Conception of \(\phi_m\) where \(1<m<M\) }
In  \eqref{eq2}, SimRIS provides g and h and \(h_{T_x-R_x}\). However, the matrix \(\phi_N\), a key matrix that optimizes the communication by maximising the SNR and is conceptualized in this paper as follows: if we assume a full knowledge of g and h and \(h_{T_x-R_x}\), we determine the phase that maximises the received signal as to be the opposite of the sum of the phases of g and h added to \(h_{T_x-R_x}\) phase. \(\phi_N\) is expressed below:

\begin{equation} \label{eq9}
\begin{aligned}
\allowdisplaybreaks
{\phi}_{m}=\diag{{\alpha}.{e^{-i\left({\phi}_{{g}_{1,m}}+{\phi}_{{h}_{1,m}}+{\phi}_{{h}_{Tx-Rx}}\right)}} & \dots & \\ {\alpha}.{e^{-i\left({\phi}_{{g}_{n,m}}+{\phi}_{{h}_{n,m}}+{\phi}_{{h}_{Tx-Rx}}\right)}}}
\end{aligned}
\end{equation}

%\begin{equation}
%\begin{split}
%\phi_m=\diag{\alpha.\exp^{-i(\phi_g_{1,m}+\phi_h_{1,m}+\phi_h_{Tx-Rx})} & \dots & \\ \alpha.\exp^{-%i(\phi_g_{n,m}+\phi_h_{n,m}+\phi_h_{Tx-Rx})}}
% \end{split}
%\label{eq9}
%\end{equation}
%Adnan, I tried fitting the matrix with resizebox and couldn't manage, could you please figure out something.

where m is the  \(m\)th intelligent surface composed of N number of elements. 
Considering \(k\) where \(1<k<N\), we have \({\phi}_{{g}_{k,m}}\) is the phase of the link between the Tx and the \(k\)th element of the RIS, 
\({\phi}_{{h}_{k,m}}\) is the phase of the link between the Rx and the \(k\)th element of the RIS and
\({\phi}_{{h}_{Tx-Rx}}\) is the phase of the direct link between the Tx and the Rx.

\subsection{Array Response Vectors and Complex Path Attenuation}
The array response vectors consist of \scalebox{0.9}{\(a\left(\phi_{c, s}^{R I S}, \theta_{c, S}^{R I S}\right), a\left(\phi_{L O S}^{R I S}, \theta_{L O S}^{R I S}\right), a\left(\phi_{R x}^{R I S}, \theta_{R x}^{R I S}\right)\)} evaluated in a similar manner for the different parameters and calculated as follows:\\
\begin{equation}
\begin{split}
a\left({\phi}_{{c}, {s}}^{{R I S}}, {\theta}_{{c}, {s}}^{{R I S}}\right)
=[1 \ldots e^{j k d\left(x \sin \theta_{c, s}^{R I S}+z \sin \phi_{c, s}^{R I S} \cos \theta_{c, s}^{R I S}\right)}\\ \ldots e^{j k d\left((\sqrt{N}-1) \sin \theta_{c, s}^{R I S}+(\sqrt{N}-1) \sin \phi_{c, s}^{R I S} \cos \theta_{c, s}^{R I S}\right)} ]^{T}
\end{split}
\label{eqarray}
\end{equation}

As for the complex path attenuations \scalebox{0.9}{\(L_{L O S}^{T-R I S}, L_{L O S}^{R I S-R}, L_{C, S}^{S I S O}, L_{C, S}^{R I S}\)},  they are evaluated in the same way using  \eqref{eqpl} considering different distances for the corresponding path \(\left(d_{T_{x}-R I S}, d_{R I S-R_{x}}, d_{c s- S I S O}, d_{c s-RIS}\right)\) where
\( d_{T_{x}-R I S}\) is distance between \(T x\) and \(R I S\) and
\( d_{R I S-R_{X}}\) represents the distance between \(\mathrm{RIS}\) and \(\mathrm{Rx}\) while \( d_{c s, S I S O}\) is the total distance Tx-Scatterer-Rx and \( d_{c s-R I S}\) is total distance Tx-Scatterer-RIS.
\begin{equation}
\label{eqpl}
\begin{split}
L_{c, s}^{R I S}=-20 \log _{10}\left(\frac{4 \pi}{\lambda}\right)\\-10 n\left(1+b\left(\frac{f-f_{0}}{f_{0}}\right)\right) \log _{10}\left(d_{c, s}\right)-X \end{split}
\end{equation}
 
 In  \eqref{eqpl},  \( X_{\sigma} \sim N\left(0, \sigma^{2}\right)\) represents a shadow factor.
As for the operating frequency \(f\), RIS operates in full band. This property of RIS is one of its biggest advantages since mmWave is the key ensuring the requirements of increasing data rates. Details of the pathloss parameters \(n\), \(\sigma\), \(f_{0}\) and \(b\) can be found in \cite{7503971}.

% No need for these after the table has been added
%We use 73 GHz in the simulations below 

%and the pathloss parameters n, \(\sigma\), \(f_{0}\) and b can be found in the pathloss parameter table in \cite{7503971}.

%there are no restrictions in frequency band choice while using RIS. As mmWave is a solution to the increasing data rate demands, 73Ghz is used in all the simulations, for which we already know the pathloss exponent n, the variance \(\sigma\) and the pathloss parameters \(f_{0}\) and b. All can be found in the Pathloss Parameter table in \cite{7503971}.

% \begin{strip}
% \begin{equation} \label{eq12}
% \begin{split}
% a\left({\phi}_{{c},{s}}^{{R I}}, {\theta}_{{c}, {s}}^{{R I S}}\right)=
% [1 \ldots e^{j k d\left(x \sin \theta_{c, s}^{R I S}+ (z \cos R) \sin \phi_{c, s}^{R I S} \cos \theta_{c, s}^{R I S}-(z \sin R) \cos \phi_{c, s}^{R I S} \cos \theta_{c, s}^{R I S}\right)} \ldots \\e^{j k d\left((\sqrt{N}-1) \sin \theta_{c, s}^{R I S}+((\sqrt{N}-1) \cos R) \sin \phi_{c, s}^{R I S} \cos \theta_{c, s}^{R I S}-((\sqrt{N}-1) \sin R) \cos \phi_{c, s}^{R I S} \cos \theta_{c, s}^{R I S}\right)} ]^{T}
% \end{split}
% \end{equation}
% \end{strip}

\begin{floatEq}
\begin{equation} \label{eq12}
\begin{aligned}
\allowdisplaybreaks
a\left({\phi}_{{c},{s}}^{{R I}}, {\theta}_{{c}, {s}}^{{R I S}}\right)=
[1 \ldots e^{j k d\left(x \sin \theta_{c, s}^{R I S}+ (z \cos R) \sin \phi_{c, s}^{R I S} \cos \theta_{c, s}^{R I S}-(z \sin R) \cos \phi_{c, s}^{R I S} \cos \theta_{c, s}^{R I S}\right)} \ldots \\e^{j k d\left((\sqrt{N}-1) \sin \theta_{c, s}^{R I S}+((\sqrt{N}-1) \cos R) \sin \phi_{c, s}^{R I S} \cos \theta_{c, s}^{R I S}-((\sqrt{N}-1) \sin R) \cos \phi_{c, s}^{R I S} \cos \theta_{c, s}^{R I S}\right)} ]^{T}
\end{aligned}
\end{equation}
\end{floatEq}

\subsection{Tilt of the Intelligent Surface}

In the tilt simulations,
 the coordinates of the RIS reflecting elements are: \(\left(x, 0, z \right)\) where x and z \({\in}\) \( {{1,{\dots}, {\sqrt{N}}-1}}\). The tilt simulation of an angle R is applied by multiplying RIS reflecting elements' coordinates by the rotational matrix. Those interfere in the RIS array response vectors \(a\left({\phi}_{c, s}^{R I S}, {\theta}_{c, S}^{R I S}\right), a\left({\phi}_{L O S}^{R I S}, {\theta}_{L O S}^{R I S}\right)\) and \(a\left({\phi}_{R x}^{R I S}, {\theta}_{R x}^{R I S}\right)\) as in \eqref{eq12}.

%$$
%\boldsymbol{a}\left(\boldsymbol{\phi}_{\boldsymbol{c}, %\boldsymbol{s}}^{\boldsymbol{R I}}, \boldsymbol{\theta}_{\boldsymbol{c}, %\boldsymbol{s}}^{\boldsymbol{R I S}}\right)=
%$$

\section{Performance of RIS-assisted Environments}\label{simulations}

In this section we conduct an evaluation of RIS-assisted radio environments based on our enhanced version of SimRIS simulator. The key simulation parameters are highlighted in \tablename~\ref{table_param}.

\begin{table}
		\caption{Key Simulation Parameters}
		\vspace{-0.3cm}
		\begin{center}
%			\rowcolors{1}{gray!25}{white}
			%\begin{tabular}{p{2.cm}p{6cm}}
			\begin{tabular}{cc}			
			
				%	\rowcolors{2}{gray!25}{white}
				\hline	
				\toprule
	%			\rowcolor{gray!50}

				\textbf{Parameter} &  \textbf{Value}    \\ \hline
			\midrule			
				No. of elements (\(N\)) & Variable     \\
		       Pathloss exponent (\(n\)) & 3.19 (non-LOS), 1.73 (LOS)
		       \\
		       Operating frequency (\(f\)) & 73 GHz
		       \\
		       Pathloss parameter (\(b\)) & 0.06 (non-LOS), 0 (LOS)
		       \\
		       Variance (\(\sigma\)) & 8.29 dB (non-LOS), 3.02 dB (LOS)
		       \\
				
				%\textbf{\textsf{X-FDR}}  & Yes & Yes & Yes & Yes \\
%				 \textcolor{blue}{Note} & \textcolor{blue}{[put a and b in mathmode throughout the table]}\\ 
				\hline
								
			\end{tabular}
		\end{center}
			\label{table_param}
	\end{table}

% \begin{center}
%  \begin{tabular}{|c | c|} 
 
%  \hline
%  Parameter & Value \\ [0.5ex] 
%  \hline \hline
%  N & Number of reflecting elements (variable) \\
%  \hline 
%  n & Pathloss exponent(3.19(non-LOS), 1.73(LOS)) \\ 
%  \hline
%  f & Operation frequency (default 73GHz) \\
%  \hline
%  b & Pathloss parameter (0.06(non-LOS), 0(LOS))   \\
%  \hline
%  \(\sigma\) & Variance(8.29dB (non-LOS), 3.02dB (LOS))   \\ [1ex] 
%  \hline
% \end{tabular}
% \end{center}

  \begin{figure}
    \centering

    \subfigure[]
    {
        \includegraphics[width=1.6in]{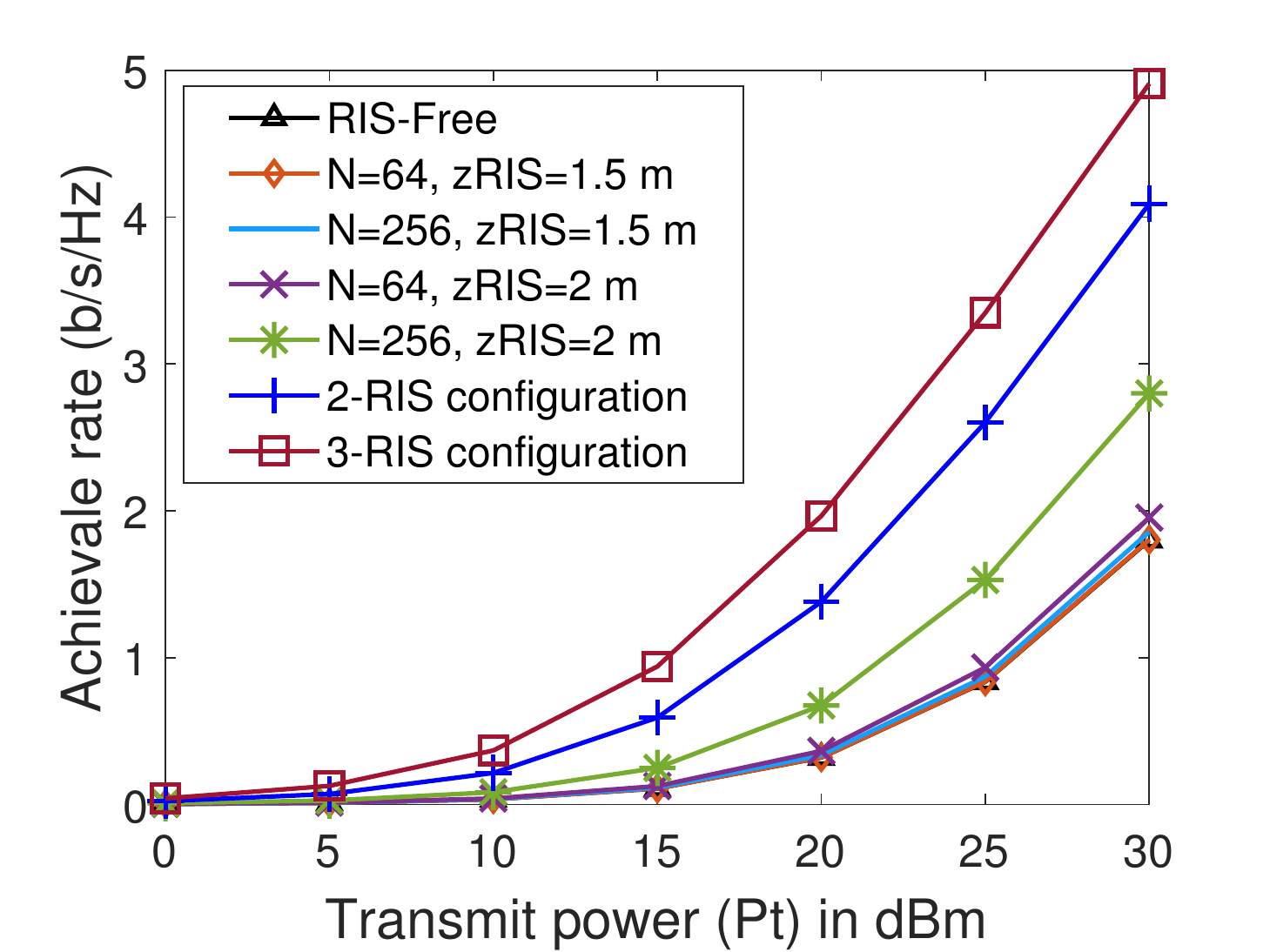}
        \label{f2a}
    }
    \subfigure[]
    {
        \includegraphics[width=1.6in]{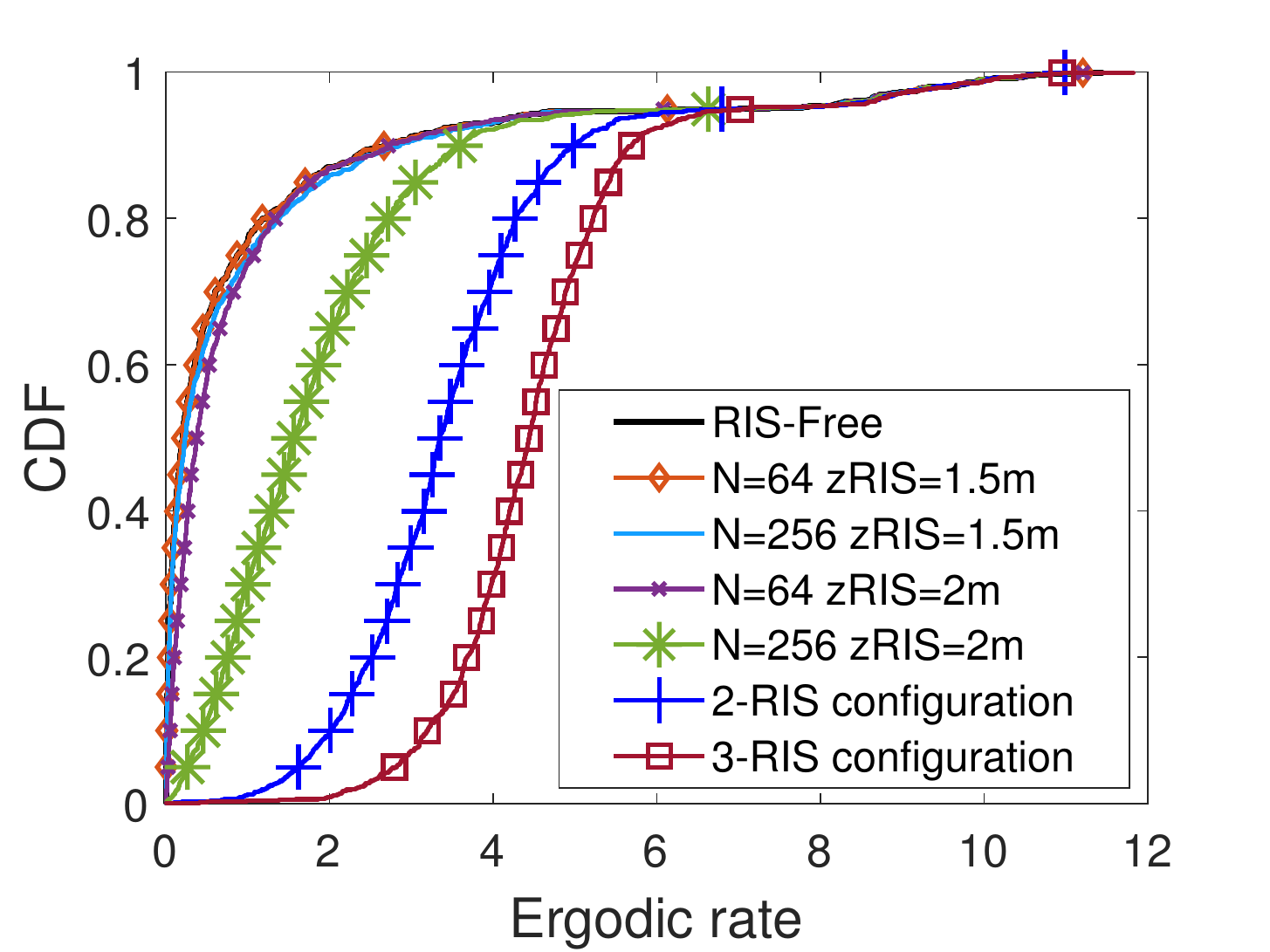}
        \label{f2b}
    }
    \caption{(a) Ergodic rate vs transmit power; (b) empirical CDF.}
    \label{f2}
    \vspace{-0.4cm}
\end{figure}

% \begin{figure}
%     \centering
%     \begin{subfigure}[b]
%         \centering
%         \includegraphics[width=4cm]{capacityAll_2.eps}
%         \caption{Achievable rate Vs PtdBm}
%%         \label{fig:y equals x}
%%     \end{subfigure}
%     \hfill
%     \begin{subfigure}[b]
 %        \centering
%         \includegraphics[width=4cm]{CDF_all.eps}
%         \caption{Empirical CDF}
%         \label{fig:three sin x}
%     \end{subfigure}
%        \caption{Three simple graphs}
%        \label{fig:three graphs}
%\end{figure}

%\section{RIS-assisted environment Simulations: }
% \begin{figure}[H]
% \begin{subfigure}{0.5\textwidth}
 %\includegraphics[width=0.9\linewidth, height=5cm]{capacityAll_2.eps}
%\caption{Caption1}
%\label{fig:subim1}
%\end{subfigure}
%\begin{subfigure}{0.5\textwidth}
% \includegraphics[width=0.9\linewidth, height=5cm]{CDF_all.eps}
%\caption{Caption 2}
%\label{fig:subim2}
%\end{subfigure}
%     \caption{(a) Capacity Vs PtdBm Indoors (b) Empirical CDF}

%%     \label{fig1}
%  \end{figure}
%\vspace{3}

A first simulation was conducted in Fig. \ref{f2}  in order to showcase the comparison of RIS-assisted SREs through three configurations (single RIS, multiple RIS and RIS-free) and multiple parameters (the number of elements in each RIS surface and the position of RIS). An indoor environment is simulated here where the Tx, the Rx, the RIS in single-assisted channel, the second RIS in 2-RIS configuration and the third RIS in 3-RIS assisted environment are located at [0, 20, 2],  [75, 35, 1], [75, 30, zRIS], [74, 30, 2], [71, 30, 2] respectively.
We notice in Fig. \ref{f2a} that RIS-free environment presents the lowest ergodic rate. The height of RIS is another crucial parameter for the ergodic rate.
The number of elements present on a single RIS is a key element for the ergodic rate. It increases with the number of reflecting elements.
Finally, the CDF in Fig. \ref{f2b} as well as the the ergodic rate plots highlight the importance of the presence of multiple RIS in a radio environment for a better communication.

\subsection{Optimal Placement of RISs} 
% \begin{figure}[H]
% \includegraphics[width=4cm]{Fig3bis_ErgodicRateinfunctionofthezpositionofRIS_1.eps}
% \includegraphics[width=4cm]{Fig4bis_ErgodicRateinfunctionofthexpositionofRIS_1.eps}
% \caption{(a) Ergodic Rate as a function of the z position of RIS (b) Ergodic Rate as a function of the x position of RIS}
% \label{f3}
%  \end{figure}
  
  \begin{figure}
    \centering
    \subfigure[]
    {
        \includegraphics[width=1.6in]{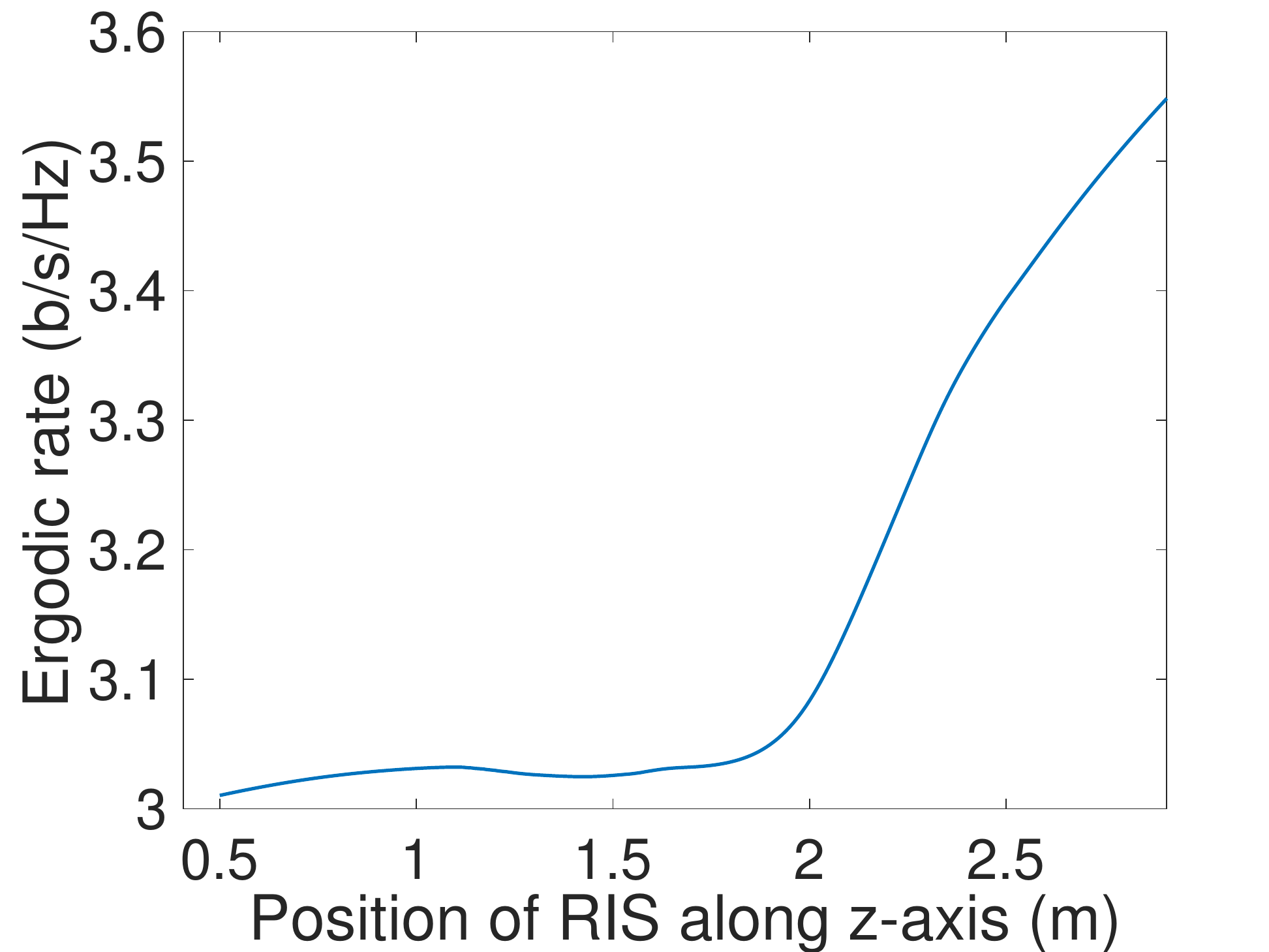}
        \label{f3a}
    }
    \subfigure[]
    {
        \includegraphics[width=1.6in]{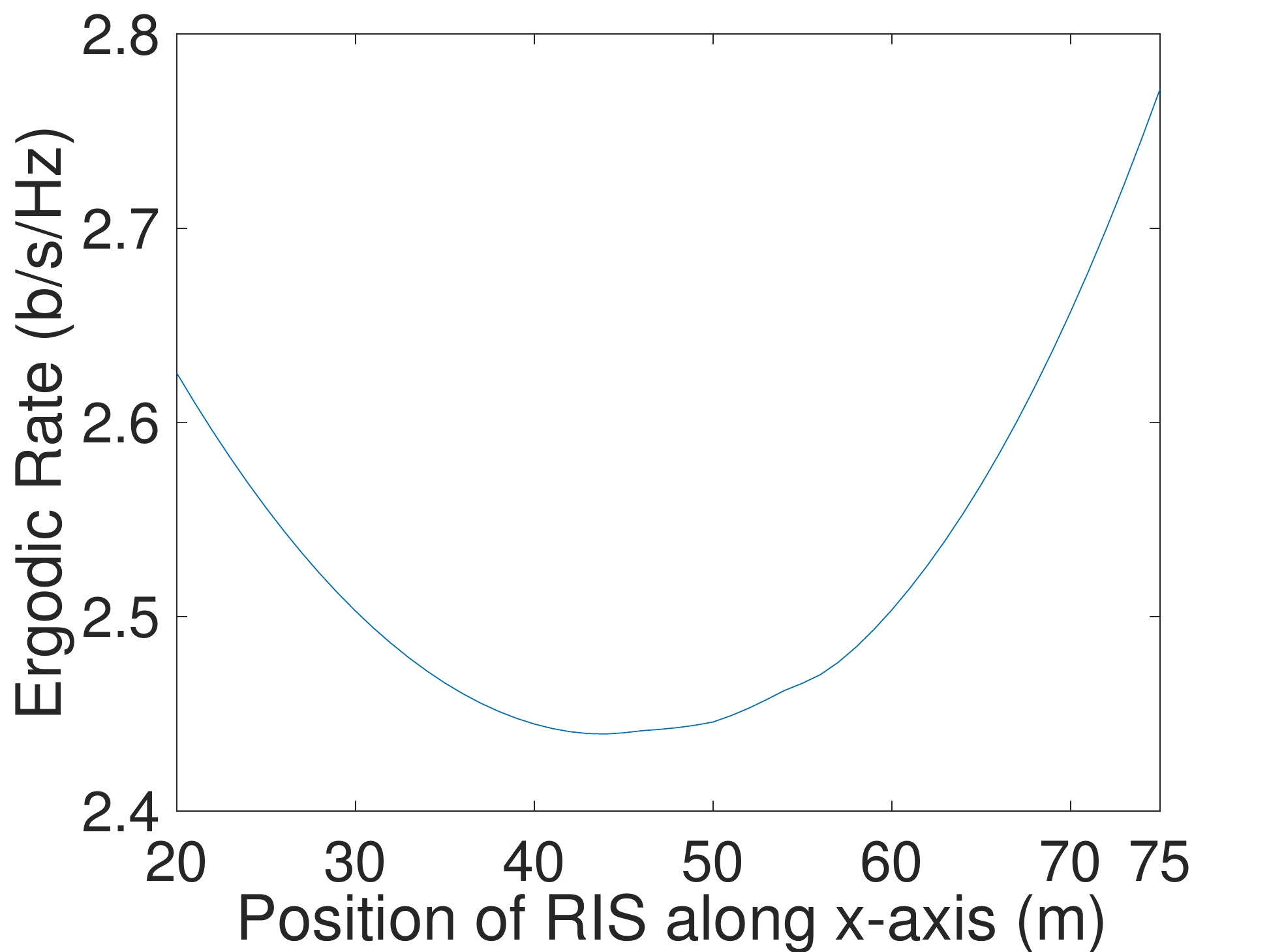}
        \label{f3b}
    }
    \caption{Ergodic rate as a function of RIS positions: (a) Ergodic Rate as a function of the z position of RIS; (b) ergodic rate as a function of the x position of RIS.}
    \label{f3}
    \vspace{-0.5cm}
\end{figure}

In this simulation, we explore optimal placement of RIS along the x-axis and z-axis. The Rx and Tx are located in their default location. In Fig. \ref{f3a}, the RIS is located at [75, 34, z] where z varies. Transmit power is fixed at 30dBm. As for the noise, it is set at -100dBm.
The ergodic rate shows very small variations while z is between 0.5 and 2m. Once the RIS is higher than 2m, which is the level of the Tx, we notice a rapid increase in the rate. That is to say, placing the RIS above Tx constitutes an optimal placement since a LOS link is applied to the Tx-RIS channel by doing so. 

%shows small and steady fluctuation from one to two meters and then a rapid increase in the rate afterwards. It is worth mentioning that the transmitter antenna is located at 2 meters in Z coordinates. 
%This could be explained with the assumption of the presence of a direct line of sight link between the transmitter and the metasurface once it reaches the level of the base station or goes beyond it. 
In Fig. \ref{f3b}, the RIS is located at [x, 30, 2] for x varying from 20m to 75m where Tx and Rx are located respectively. Fig. \ref{f3b} shows that the nearer the RIS gets either to the Rx or to the Tx, the rate rate increases. This is explained by the fact that the probability of the presence of a LOS link becomes higher in the RIS-Rx channel (when RIS is at the Rx side) or  in the Tx-RIS channel (when RIS is at the Tx side),  which means more significant received power. As a result, the ergodic rate increases considerably.
%It gives an overview on the optimal placing of the intelligent surface. 

%It sho closer to the endpoints (receiver and with this effectiveness near the transmitter)
%The RIS is at a location fixed at  where x varies between 25 and 70. It is also worth mentioning that the transmitter is on 20m in X-coordinates and the receiver is at 70m in x-Coordinates. 
%we can see that the closer we get to the base station where the receiving antenna is located, the higher the capacity becomes. From the receiver side, when the RIS is closer, the probability of the direct line of sight from RIS to the receiver is higher, therefore, the received power is more significant. As at the base station end, the impinging wave is more likely to be from a LOS path between TX and the metasurface which means the power of the reflected wave becomes higher when the RIS is closer (around 20m in x-coordinates). 

\begin{figure}
 \includegraphics[width=0.32\textwidth, center]{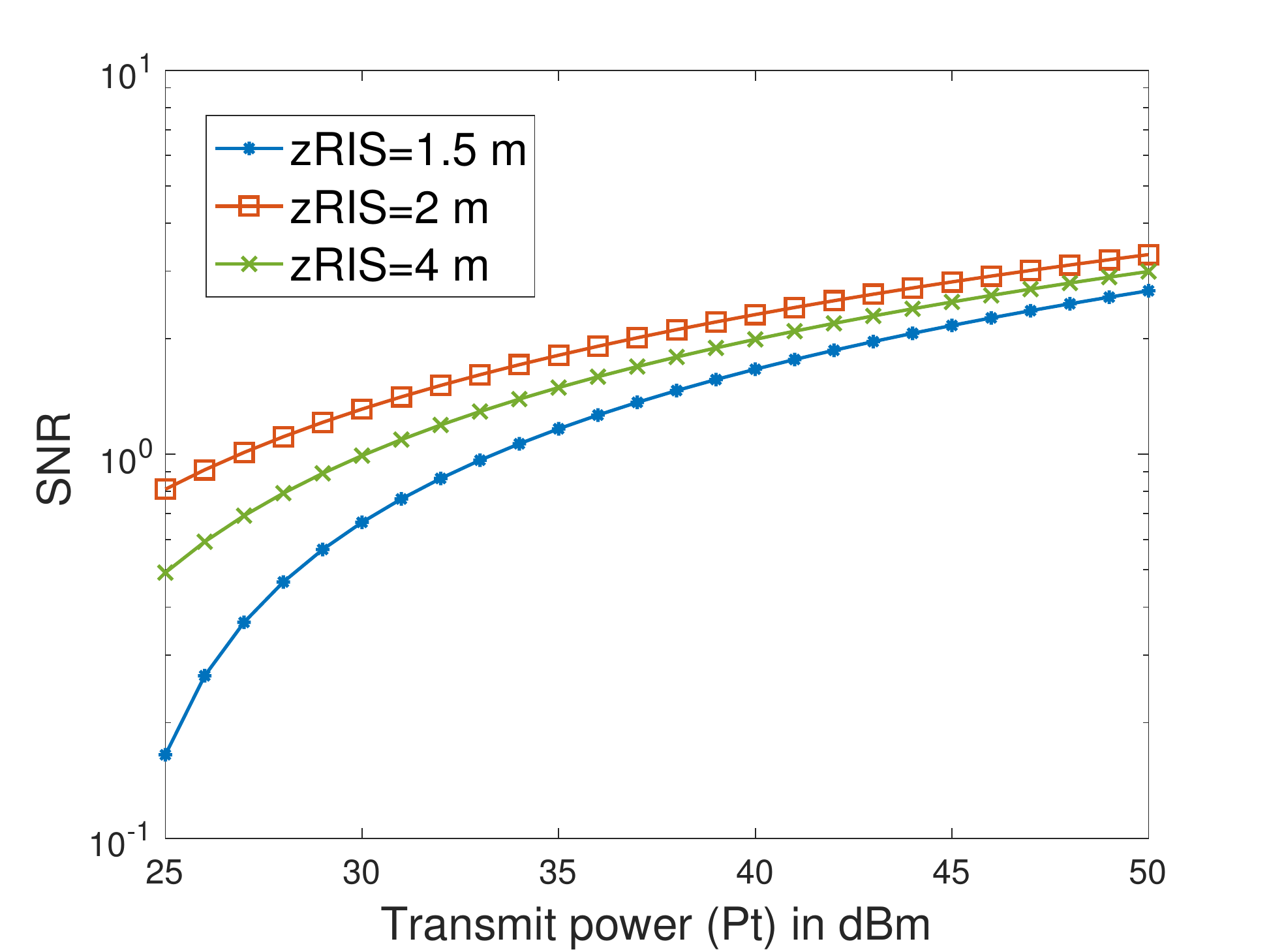}
     \caption{Signal-to-noise ratio (SNR) in an RIS-assisted channel for different RIS height levels (zRIS).}
     \label{f4}
%     \label{fig1}
\vspace{-0.4cm}
  \end{figure}
  
\subsubsection{Reliability Enhancement through RIS positioning}

While processing the (received) SNR for a simulated communication channel assisted by an RIS located in [75, 30, z] for different z \( \in( 2, 3, 4)\) in Fig. \ref{f4}, a considerable increase of SNR with the increase of the level of the RIS is noticeable.
Also, for smaller values of transmit power, the gap in SNR between the sets of z-coordinates of RIS for different sets is wider. However, the value of the SNR gets higher for more powerful transmit power.
In short, not only placing the RIS higher at or above the Tx antenna level increases the ergodic achievable rate, but it also enhances the reliability of the communication network.
\subsection{Tilt and Rotation of RIS}
%  \begin{figure}[H] 
% \includegraphics[width=4cm]{Fig10bis_CapacityVsRISXaxisrotationRISinxzplan.eps}
% \includegraphics[width=4cm]{Fig11bis_CapacityVsRISYaxisrotationRISinyzplan.eps}

%     \caption{(a) Ergodic rate as a function of RIS X-axis rotation where RIS in XZ plane initially (b) %Ergodic rate as a function of RIS Y-axis rotation where RIS in YZ plane initially}
%    \label{f8}
%  \end{figure}

\begin{figure}
    \centering

    \subfigure[]
    {
        \includegraphics[width=1.6in]{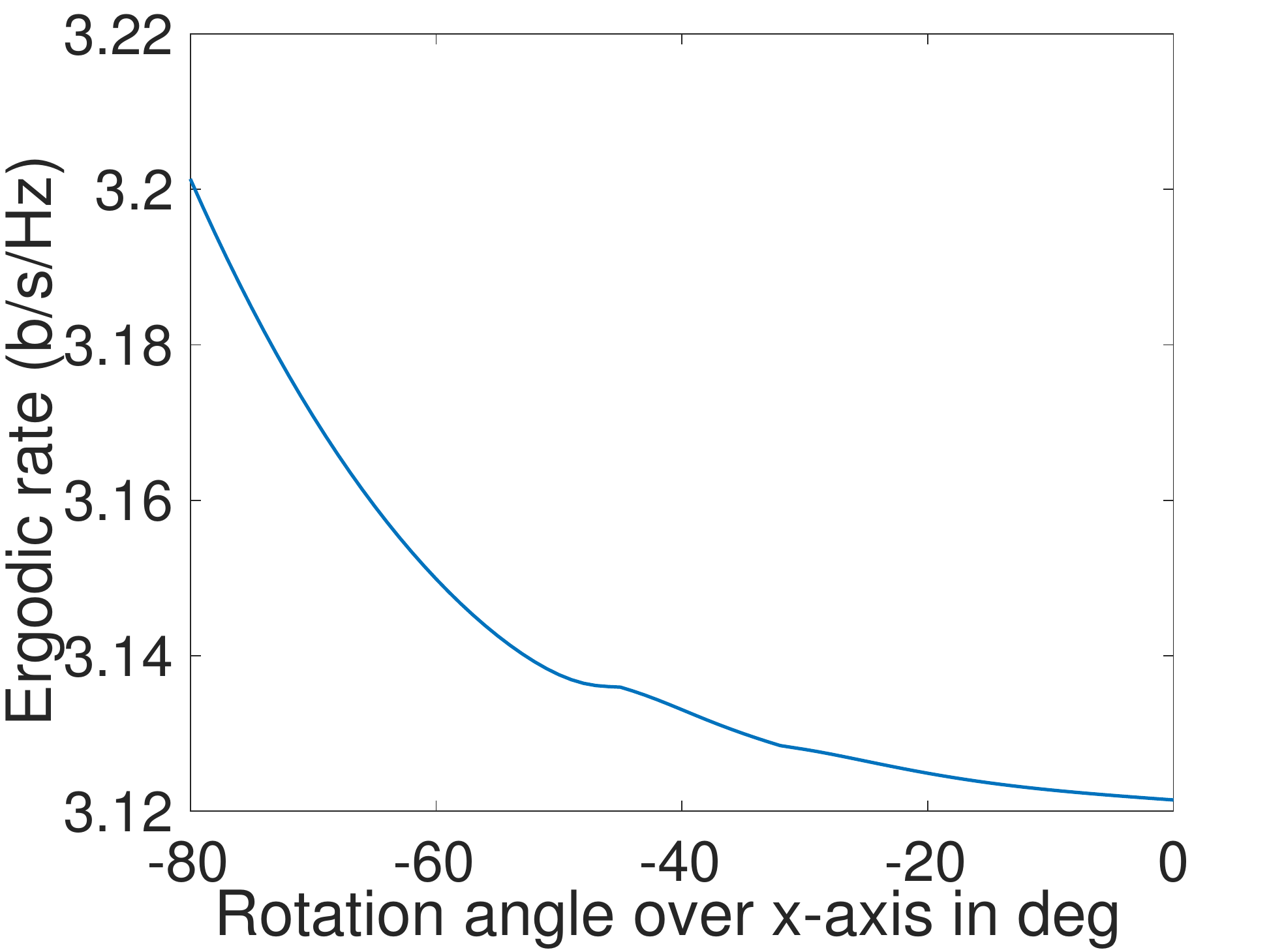}
       \label{f5a}
    }
    \subfigure[]
    {
        \includegraphics[width=1.6in]{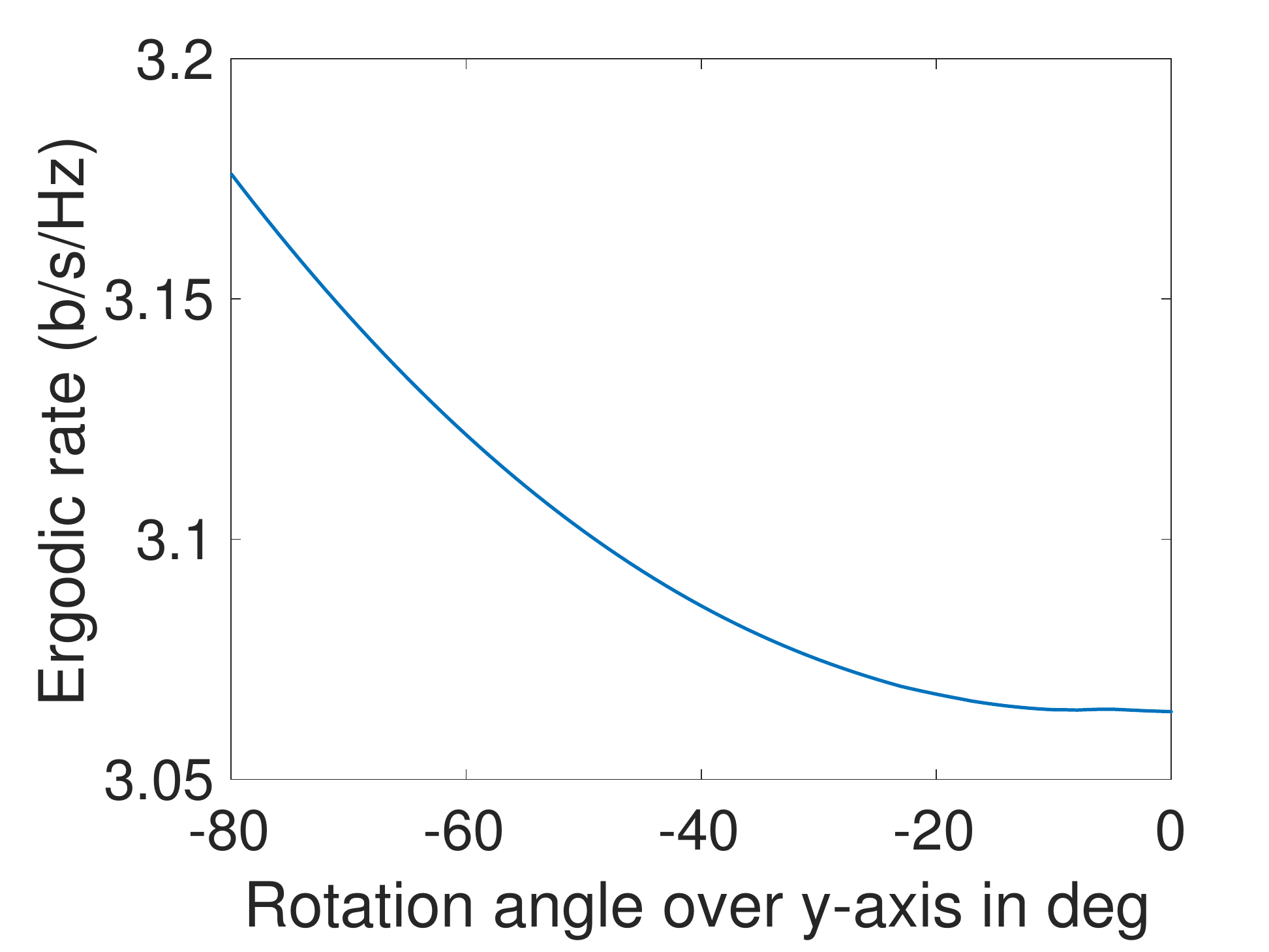}
        \label{f5b}
    }
    \caption{(a) Ergodic rate as a function of RIS x-axis rotation where RIS in xz-plane initially; (b) ergodic rate as a function of RIS y-axis rotation where RIS in yz-plane initially.}
    \label{f5}
    \vspace{-0.5cm}
\end{figure}
  
%  \begin{figure}
 %\includegraphics[width=9cm]{Fig11bis_CapacityVsRISYaxisrotationRISinyzplan.eps}

  %   \caption{}

  %\end{figure}
We prove in this section that an RIS that focuses the reflected signal towards the Rx increases the ergodic rate. We apply 15 dBm transmit power and -100 dBm as noise power. Initially, the RIS is located at [70, 30, 2]  at a 0° tilt angle on xz-plane. In Fig. \ref{f5a}, we plot the ergodic rate while we tilt the surface over x-axis to point towards the Rx which is located  [70, 35, 1] (towards -80°). We repeat the same process in Fig. \ref{f5b} with RIS located at [75,  35,  2] on yz-plane initially and rotating over y-axis to point towards the Rx.

%Major characteristic of RIS is controlling the surrounding environment. It makes it a deterministic and no longer a probabilistic hazardous environment. A new simulation idea emerged where the meta-surface tilts towards the receiver. To this end, our simulation in this section aims at exploring the Ergodic rate as a function of the rotation of the metasurface.  Received power and noise power equal 15 dBm and -100 dBm respectively. 
%Initial state of RIS is on XZ plane with first element of the metasurface is at [70, 30, 2] for Fig. \ref{f8}a (rotation angle is null) and tilting over X-axis pointing towards the receiver end which is positioned at [70, 35, 1] until reaching rotation angle of .  The transmitter coordinates are [0, 32, 2].
%Same process goes for Fig. \ref{f8}b where this time RIS is on YZ plane, initially with first element on [75, 32, 2], the RIS rotates over Y-axis with all its elements to point down towards the receiver which is positioned at [70, 35, 1]. The transmitter is positioned at [0, 30, 2].

Both of the simulations show an increase in the ergodic rate the moment we start tilting the RIS to point towards the Rx (in the order of 0.08 and 0.12 b/s/Hz for -80° tilt angle). In short, this simulation proves that RIS performs better when it focuses the beam towards the Rx.

\subsection{Number of Reflecting Elements on an RIS}
\begin{figure}
 \includegraphics[width=0.32\textwidth, center]{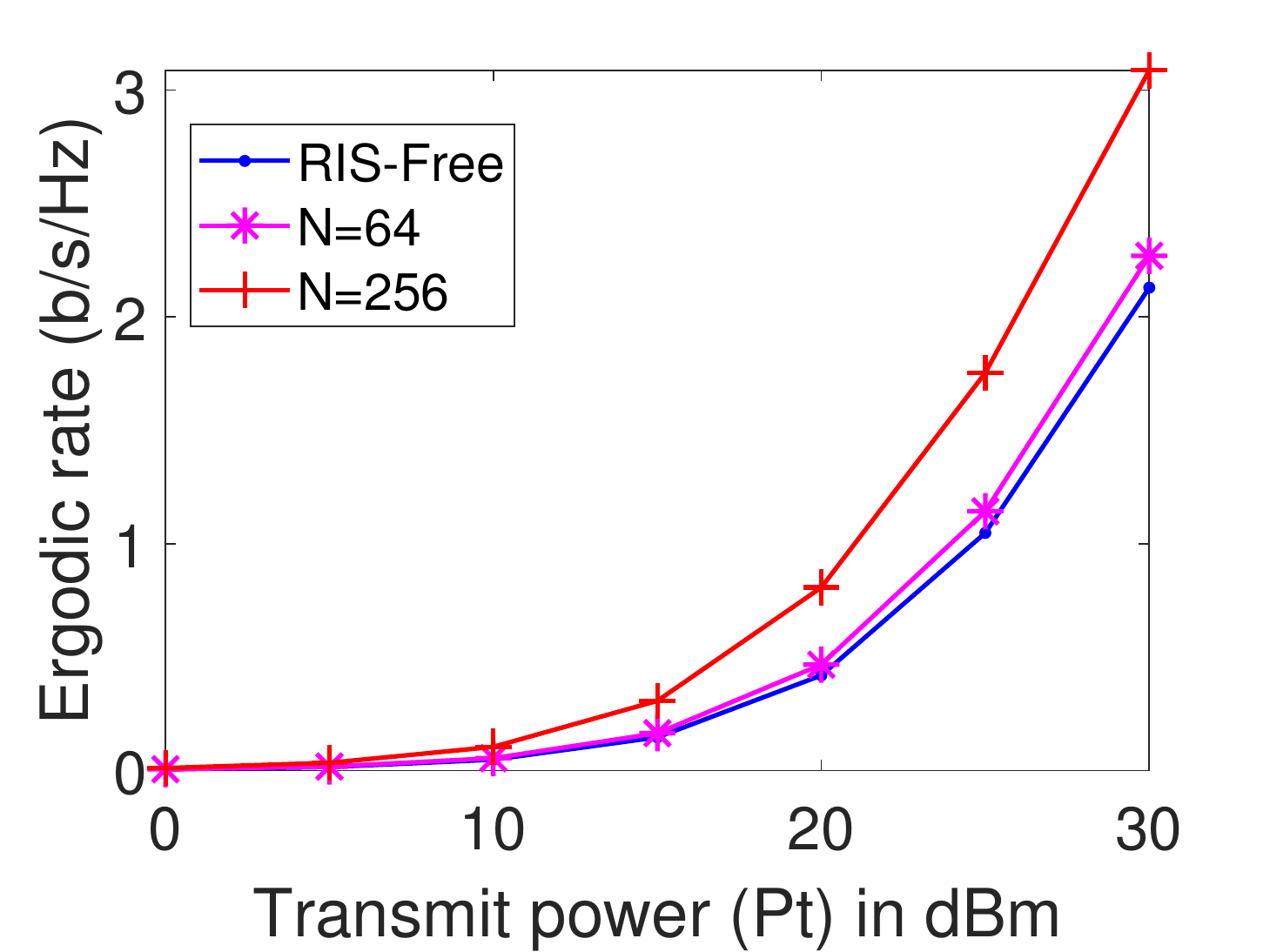}
     \caption{Influence of number of elements \((N)\) on ergodic rate in  a single-RIS configuration.}
\label{f6}
\vspace{-0.5cm}
 \end{figure}
 
The reflecting elements on a single RIS allow the control of more impinging waves. At the Rx end, the received power is proportional to the square of the number of reflecting elements N which represents the number of the independently controlled phases from a single surface. The equation is as follows \cite{basar2019wireless}:
\begin{equation}
P_r \approx (N+1)^2 P_t (\frac{\lambda}{4\pi d})^2,
\end{equation}
where \(\lambda\) is the wavelength, \(d\) is the distance Tx-RIS-Rx and \(P_t\) is the transmit power. The following simulations show how both the capacity and reliability increase with N. Three plots corresponding to 3 values of N are shown in Fig. \ref{f6}: N=64 in purple, 256 in red and in blue, there is RIS-free environment. Two results can be deduced from it. First, the increasing capacity with the increasing N confirms what has been dictated in the theory. Second, the 64-element RIS does not bring major improvements to the environment.
\subsection{Optimal RIS Placement in a Multi-RIS Environment } 
  \begin{figure}
    \centering

    \subfigure[]
    {
        \includegraphics[width=1.6in]{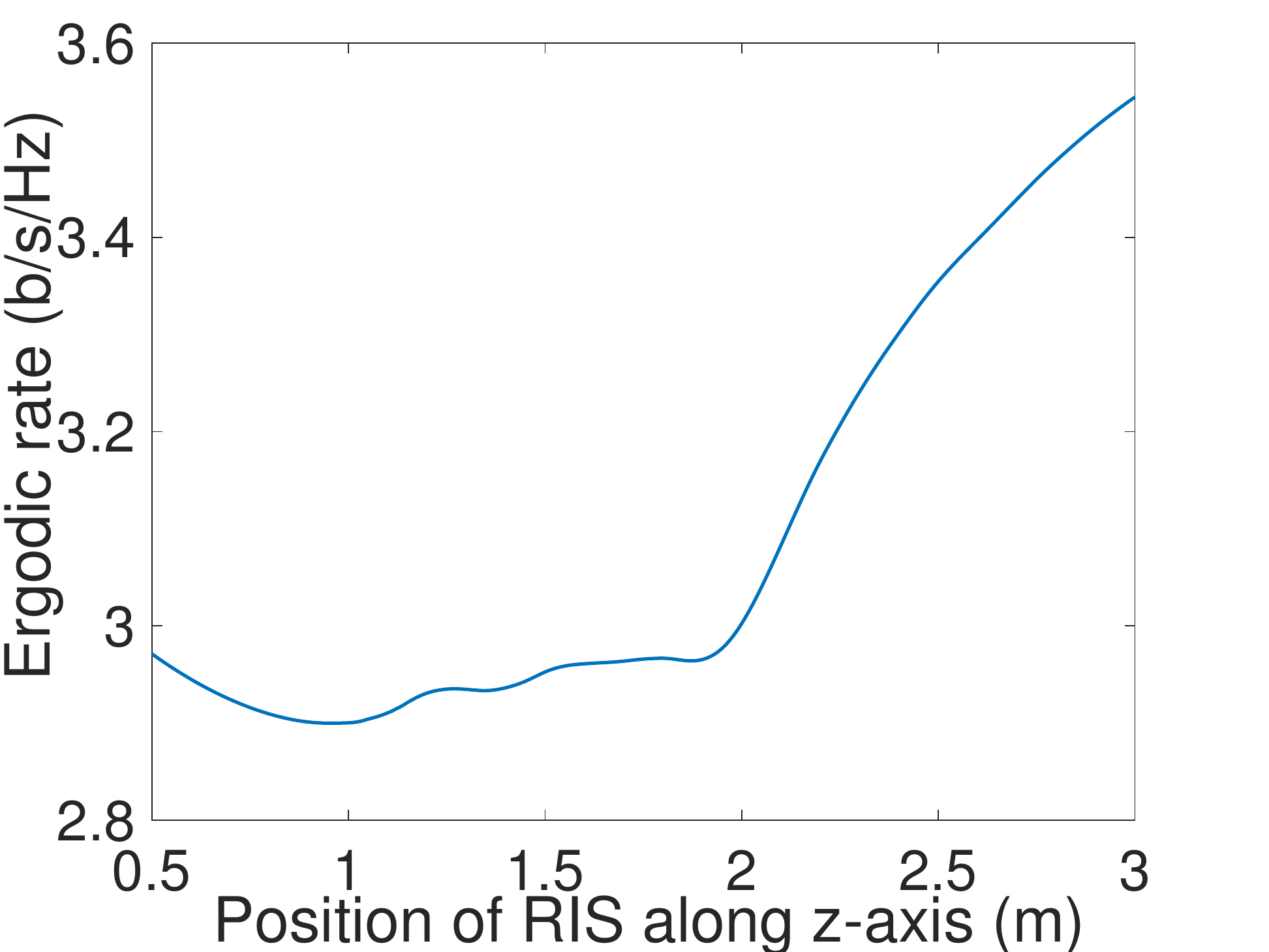}
        \label{f7a}
    }
    \subfigure[]
    {
        \includegraphics[width=1.6in]{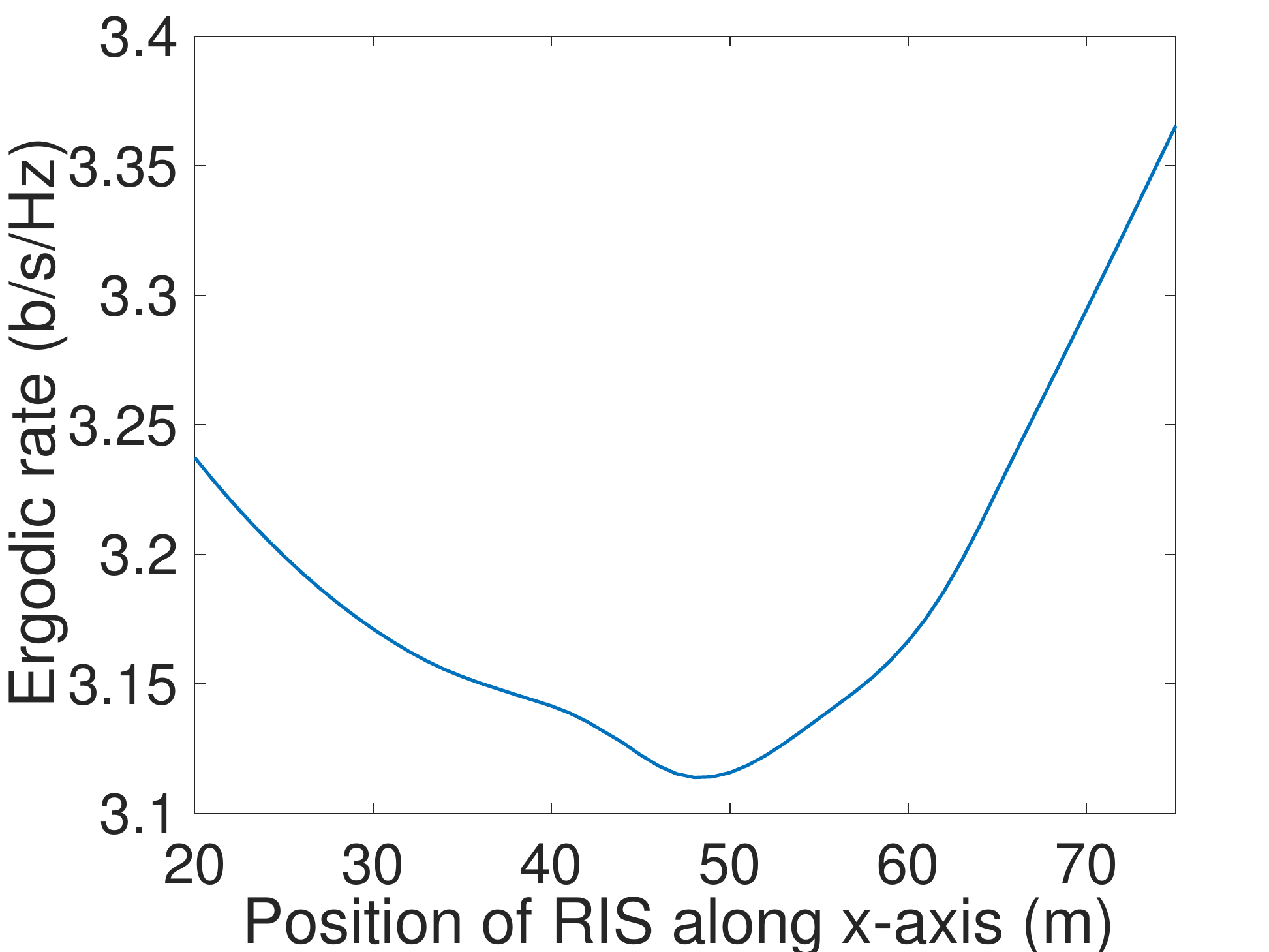}
        \label{f7b}
    }
    \caption{Ergodic rate in a multiple RIS assisted environment: (a) Ergodic rate as a function of z-coordinates; (b) ergodic rate as a function of x-coordinates.}
    \label{f7}
    \vspace{-0.5cm}
\end{figure}  
We explore here the influence of multiple RIS configuration on the performance of the communication channel as a function of varying location of one RIS [x, 34, z] while fixing the location of the other one at [75, 30, 2]. In fig. \ref{f7a}, x varies between 25m and 75m and z fixed at 2m. In fig. \ref{f7b}, z varies between 1m and 3m.  As a result, plots held the same variation as the ones obtained in the previous section. Therefore, we conclude that ergodic rate increases when RIS gets closer to the base stations.  Moreover, we achieve higher data rates when we implement RIS above the level of Tx (z=2m) because the probability of a LOS from Tx to the intelligent surface is higher. Furthermore, and most importantly, this simulation shows that we can achieve an additional 1 b/s/Hz gain in ergodic rate in a multiple-RIS configuration compared to single-RIS configuration.

%The position of RIS in space has a big importance in order to get desirable outcome. In the simulation Fig. \ref{f4}, we are exploring the benefits of having multiple RISs in function of X and Z coordinates. For that reason, we set one RIS on a fixed position: [75, 30, 2]. The coordinates of the second RIS on the other hand are: [x, 34, z], where for the first simulation z varies from 1 to 3 meters and x = 75 represented in Fig. \ref{f4} a. As for the second simulation, x varies from 20 to 70 and z = 2 in Fig. \ref{f4}(b). 
%Compared to the results obtained in the first section of this paper we can clearly see that while fixing the position of one RIS and translating the second RIS whether on Z axis or X axis, the rate variation remains the same. Meaning, the closer the RIS is to the antennas the higher is the rate. Furthermore, the higher we implement the RIS the more likely to have a direct line of sight from the transmitter to the 
%metasurface. On the other hand, what is worth of mentioning is the considerable gain in the rate of more than 1 bit per second per Hertz when implementing another RIS. \hfill \break 

\subsection{Impact of Number of Elements in a Multi-RIS Environment}

\begin{figure}
    \centering

    \subfigure[]
    {
        \includegraphics[width=1.6in]{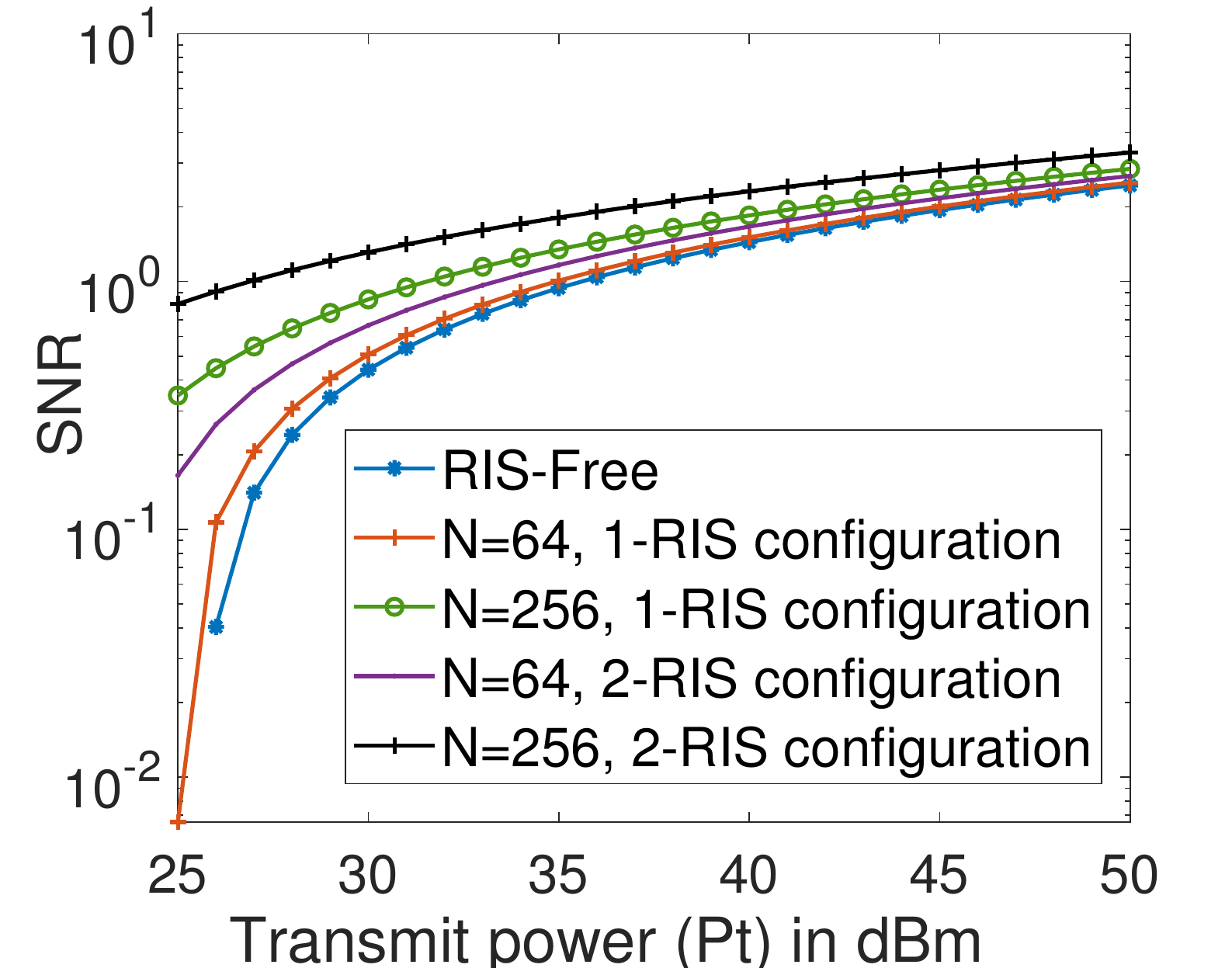}
        \label{f8a}
    }
    \subfigure[]
    {
        \includegraphics[width=1.6in]{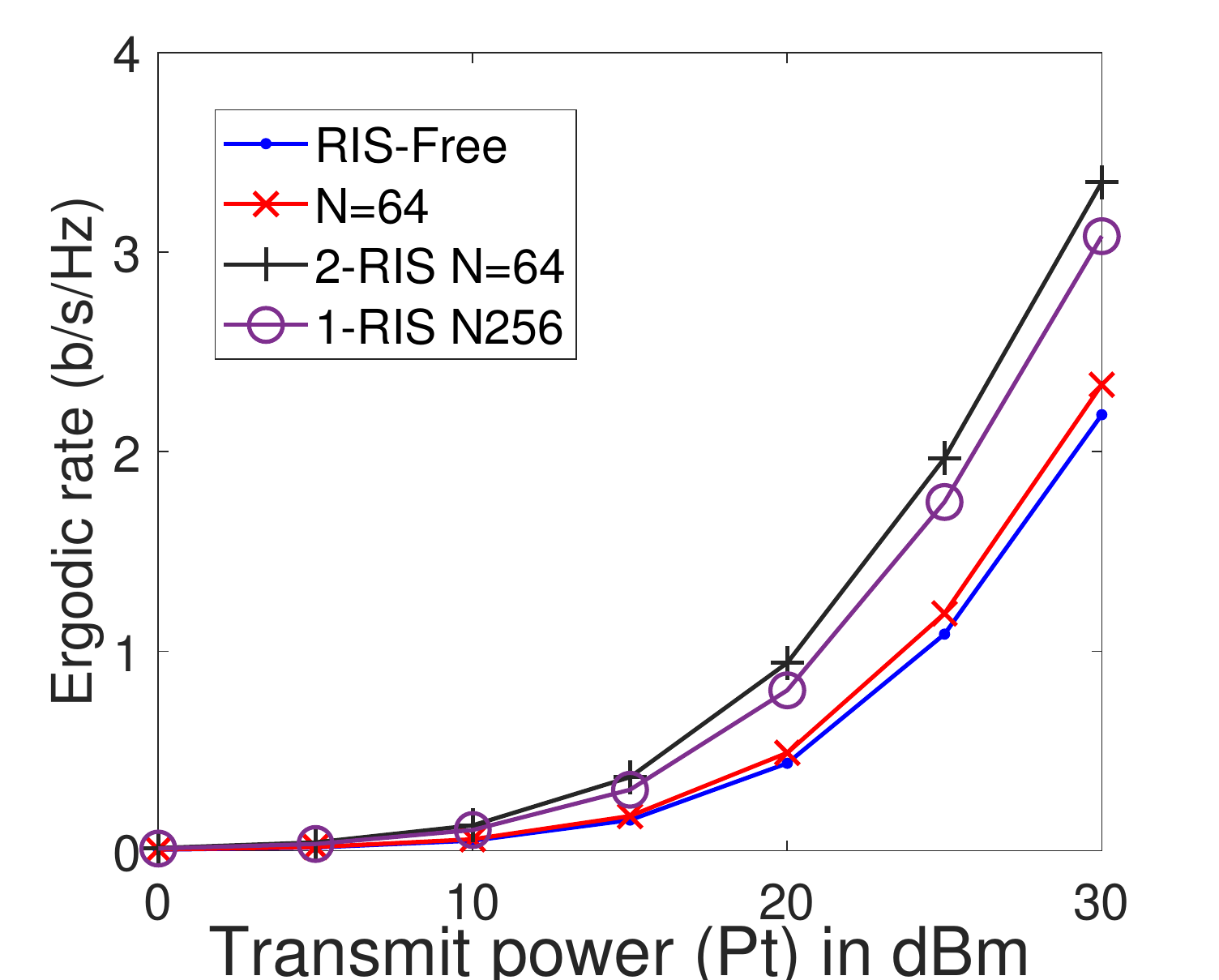}
        \label{f8b}
    }
    \caption{Influence of number of elements \((N)\) in a multi-RIS configuration: (a) SNR; (b) ergodic rate.}
    \label{f8}
    \vspace{-0.5cm}
\end{figure}
 
The SNR simulation in Fig. \ref{f8a} confirms the benefits of multiple RIS deployment and the increase of N. The reliability of the communication environment is enhanced with more RISs and a higher number of elements N on each surface. We notice a 0.4 dB SNR gain in 1-RIS configuration against 0.87 dB SNR gain in a multi-RIS configuration compared to RIS-free environment for a \(P_t\) = 26 dBm.  
On the other hand, in the 64-element RIS case, it is no longer overlapping with RIS-free environment for multi-RIS configuration in Fig. \ref{f8b}. It also demonstrates similar performance as the 2-RIS configuration which increased the ergodic rate by 0.8 b/s/Hz for \(P_t\) = 30 dBm. This highlights a trade-off consisting of whether to use a single RIS composed of 256 reflecting elements or 2-64 element RIS. It is a trade-off between deploying many RISs of smaller number of reflecting elements or deploying fewer RISs with higher number of elements. 

\subsection{Multiple Users in an RIS-assisted Environment}

\begin{figure}
 \includegraphics[width=0.35\textwidth, center]{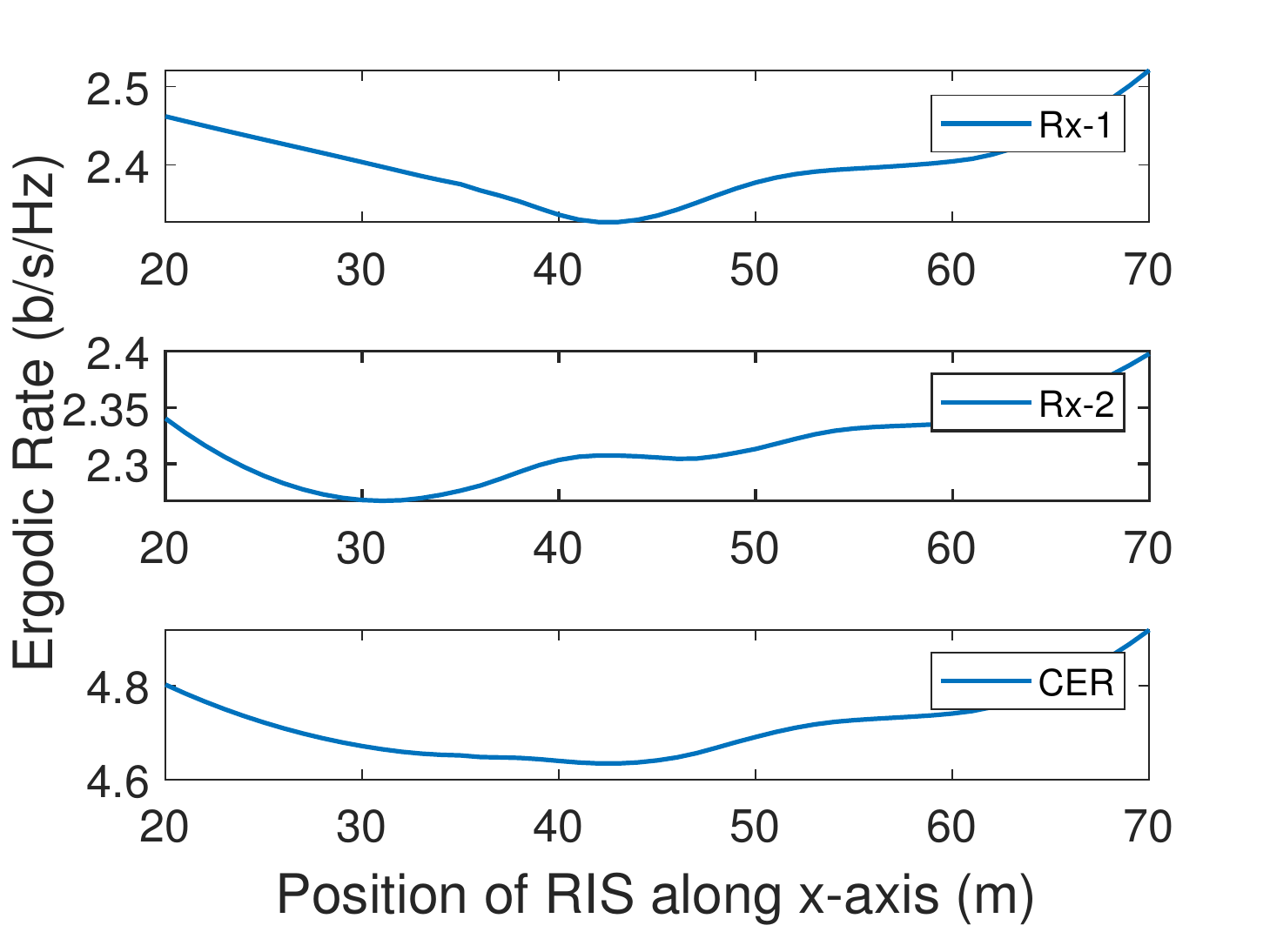}
     \caption{Ergodic rate at user 1 and 2 (Rx-1, Rx-2) and cumulative ergodic rate (CER) in a 256-element RIS-assisted SRE where 128 elements are allocated for each user. }
\label{f9}
\vspace{-0.5cm}
  \end{figure}
RIS is envisioned within a communication platform that not only serves one but multiple users. In this section, we investigate the cumulative ergodic rate in a 2-user scenario through a single RIS assisted environment. The RIS consists of 256 controllable reflecting elements. In this simulation, we split those elements to serve the two users Rx-1 and Rx-2 positioned at [70, 32, 1] and [70, 35, 1] respectively. We allocate 128 reflecting elements of the RIS located at [x, 30, 2] for each user. We then plot in Fig. \ref{f9} the ergodic rate for each user as a function x that varying between 20 and 70. 
First thing to notice is that the closer the RIS gets to the users or the base station, the ergodic rate increases. 
On the other hand, the two users are two metres apart in the y-axis, which slightly impacts capacity performance with a 0.125 b/s/Hz difference between the users. Nevertheless, the ergodic rate is much higher for both users by allocating a certain number of elements to the users by only using a single RIS compared to the ergodic rate in RIS-free environment present in Fig. \ref{f2a} where for 30 dBm transmit power, the ergodic rate is less than 2 b/s/Hz. This opens up the door towards future research opportunities that covers an optimal reflecting elements' allocation that could maintain steady and even performance for the users.   

% you mentioned f3a in above paragraph which was a serious mistake!

\section{Key Insights and Concluding Remarks}\label{conclusion}

RIS is a promising technology for 6G wireless systems. It was the aim of this paper to conduct a holistic evaluation of RIS-assisted SREs. We have conducted a simulation-drive performance evaluation under different scenarios and configurations. 
The key findings and insights on system performance are highlighted as follows. 

\begin{itemize}
\item Optimal RIS placement is next to the endpoints (Tx or Rx) and above Tx level where the capacity increases by 0.3 b/s/Hz near the Rx and by 0.2 b/s/Hz by the Tx. SNR also increases from 0.165 dB in case for 1.5m RIS elevation from the ground to 0.810 dB for 4m elevation, scoring a gain of 0.645 dB.
Moreover, the ergodic rate in an RIS- assisted SRE increases with the  number of reflecting elements present on a single RIS. It increased from 2.1 b/s/Hz in an RIS-free environment to 2.3 b/s/Hz for 64-element and 3.1 b/s/Hz for 256-element RISs.  

\item Multiple RIS-assisted smart radio environment drastically improves capacity and reliability. It has shown an increase of 1 b/s/Hz in ergodic rate and 0.46 dB in SNR for a 256 element 2-RIS configuration compared to a single-RIS environment for 25 dBm transmit power. This multiple RIS configuration has also proven its effectiveness for a 64-element RIS. As a matter of fact, ergodic rate augmented from 2.3 b/s/Hz in a single 64-element RIS (a slightly improved communication environment) to 3.3 b/s/Hz in a 2-RIS configuration (a stronger performance similar to single 256-element RIS configuration).
\item The tilt (rotations over x-axis and y-axis) of the RIS was also investigated. We found that directing the beam to point to the receiver by tilting the RIS increased the capacity by 0.11 b/s/Hz.
\item In the case of multiple users in a 256-element RIS assisted environment, we split the reflecting elements from the RIS to serve both users by allocating 128 elements for each. We noticed a better performance with an increased ergodic rate for both users with a slight 0.12 b/s/Hz difference between the two.  
\end{itemize}

Key future work directions include optimal design of an SRE with a single RIS having sufficiently large number of elements or multiple RISs with relatively small number of elements and adaptive allocation of elements to each receiver in multi-user environments. 

%Future work:
%Our simulations for the number of elements highlighted a trade-off in the RIS conception and manufacturing, i.e., whether building one RIS of 256  elements or two RISs comprising 64 elements each. The future work willl investigate the optimal design for a SRE. Such investigation is also important from economic perspective as it dictates the cost of overall system. 

%A topic that we believe should be investigated in future work since it could reduce the cost of RIS manufacturing and assuring same performance.  

%Another important aspect is the allocation of the number of reflecting elements of an RIS to  multiple receivers. Finding an adaptive allocation for each user is crucial to improving system-level reliability and capacity. 

%finding an optimal allocation that assures equal reliability and capacity for all users. \\
%Key simulation parameters include zRIS: RIS position on z-axis, N: number of reflecting elements on a single RIS, Rx-1 and Rx-2: users 1 and 2, CER: cumulative ergodic rate. 
%It is only dutiful to summarize the key points made on this report to %conclude this paper.

%\begin{figure}
%\centerline{\includegraphics[width=8.5cm]{MacModule.pdf}}
%\caption{Transmission process at the eNodeB.}
%\label{fig1}
%\vspace{-0.5cm}
%\end{figure}

%\vspace{12pt}
\bibliographystyle{IEEEtran}
\bibliography{bibliography.bib}
\end{document}